\RequirePackage{lineno}
\documentclass[aps,prb,twocolumn,superscriptaddress]{revtex4}

\usepackage{graphicx}
\usepackage{bm}
\usepackage{amsmath}
\usepackage{amssymb}
\usepackage{hyperref}

\begin{document}

\title{Periodic ripples in suspended graphene}

\author{Zhao Wang}
\affiliation{LITEN, CEA-Grenoble, 17 rue des Martyrs, 38054 Grenoble Cedex 9, France}

\author{Michel Devel}
\affiliation{Institut FEMTO-ST, Universit\'{e} de Franche-Comt\'{e}, CNRS, ENSMM, UTBM, 26 chemin de l'\'{e}pitaphe, F-25030 Besan\c{c}on Cedex, France}

\begin{abstract}
We study the mechanism of wrinkling of suspended graphene, by means of atomistic simulations. We argue that the structural instability under edge compression is the essential physical reason for the formation of periodic ripples in graphene. The ripple wavelength and out-of-plane amplitude are found to obey 1/4-power scaling laws with respect to edge compression. Our results also show that parallel displacement of the clamped boundaries can induce periodic ripples, with oscillation amplitude roughly proportional to the 1/4 power of edge displacement. The results are fundamental to graphene's applications in electronics.
\end{abstract}

\maketitle
Graphene's unique electronic properties makes it ideal candidate for integrated circuits component. Theoretically, conduction in a perfectly flat graphene sheet can be ballistic, despite what observed experimentally is quite different. This is because that ripples are manifested\cite{Vazquez2008,Diaye2006,Lui2009} by developing a band gap, introducing additional effective magnetic fields.\cite{CastroNeto2009} Hence, understanding the rippling mechanisms is crucial for applications of graphene in nanoelectronics. Recently, it is reported that the ripple structure can be controlled by thermal treatment.\cite{Bao2009} This brings out a straightforward way to the band gap engineering of graphene.\cite{Miranda2009} Some relevant theoretical works have attempted to study this rippling using molecular dynamics (MD), simulating suspended graphene under axial compression at different temperatures.\cite{Abedpour2010} Graphene rippling by thermal treatment is known to be related to its negative thermal expansion coefficient (TEC),\cite{Mounet2005,Abedpour2010} and its membrane-natured mechanical properties.\cite{Lee2008,Cerda2003} However, the intrinsic mechanism responsible for the periodic rippling of suspended graphene is not fully understood.

\begin{figure}[ht]
\centerline{\includegraphics[width=9cm]{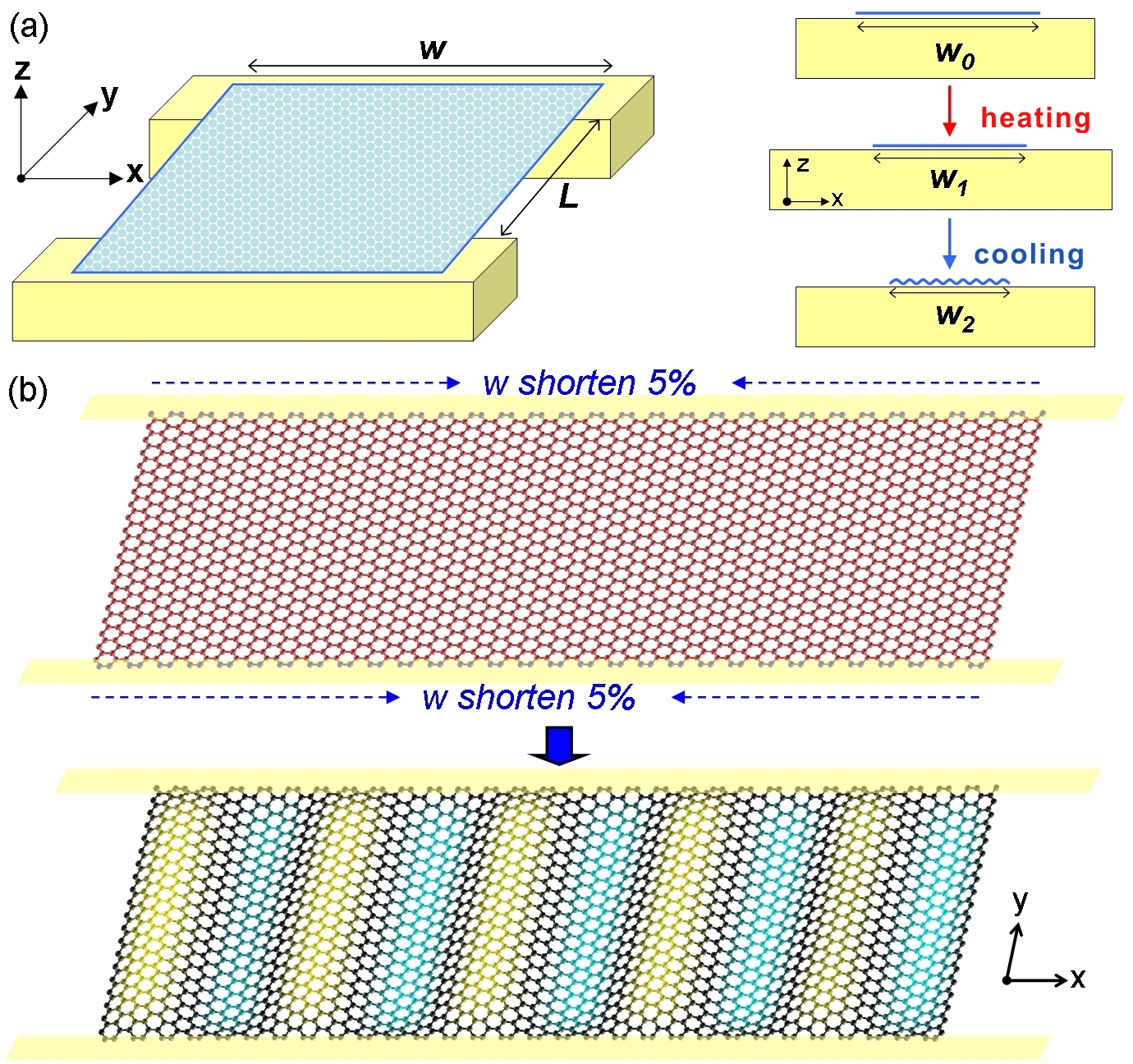}}
\caption{\label{fig:Schema}
(Color online) (a) Schematic of edge compression of suspended graphene after heating and cooling. (b) Topographic diagram of edge-compression-induced periodic ripples in a graphene sheet ($w \times L = 13 nm \times 5 nm$). Color scale shows the height profile (position along $z$ axis).} 
\end{figure}

From a mechanical point of view, there are in general two ways to induce periodic ripples in a suspended thin film: \textit{1} Stretching in the axial direction which is perpendicular to the fixed boundaries, or, \textit{2} Compression of the fixed edges in the lateral direction.\cite{Timoshenko1963} Here we focus on the later case (so-called edge contraction), in view of the experimental observation of Bao \textit{et al.}, which shows biaxial compression of graphene after annealing.\cite{Bao2009} We show the origin of edge compression due to heat treatment in Fig.\ref{fig:Schema} (a): During heating, a difference in thermal deformation is created between the suspended graphene and the substrates, due to their different TEC. The graphene is stretched by the friction force given by the expanding substrate during the heating process. When, on the contrary, the system is cooling down, compressive force will be applied to the fixed boundaries from the interface. The deformation of the graphene boundaries becomes irreversible, because graphene exhibits high structure instability under in-plane compression.\cite{Shenoy2008} 

In our MD simulation, we start with a graphene sheet suspended between two parallel supports, with an in-plane compressive strain imposed on its two edges fixed on substrate. The equations of motion are integrated by the Verlet algorithm with a time step of $1$ fs. The system consists of $1000-20000$ atoms. A Nos\'{e}-Hoover thermostat is used to help the system to reach equilibrium at $300$ K in the first 500000 simulation steps. We let the system progressively reach further equilibrium in the next 500000 steps without manual thermal control. Further details about the simulation techniques can be found elsewhere.\cite{zhao2010ss} We note that the compressive strain is applied on the fixed edges along the direction parallel to the trench, in order to generate ripples in experimentally observed direction. This makes the rippling orientation studied in this work perpendicular to that simulated in the relevant work of Abedpour \textit{et al.}.\cite{Abedpour2010} We also note that the here-simulated ripples are different from the intrinsic ones due to thermal fluctuation, which were found to distribute randomly over the surface with height variation down to the atomic level.\cite{Fasolino2007}

The total interatomic potential involves many-body terms,\cite{Stuart2000a} as a collection of that of individual bonds,

\begin{equation}
\label{eq:1}
U^p=\frac{1}{2}\sum\limits_{i=1}^N{\sum\limits_{\substack{j=1 \\ j\ne i} }^N{\left[ 
\begin{array}{l}
\varphi^R\left(r_{i,j}\right)
-b_{i,j}\varphi^A\left(r_{i,j}\right)
+\varphi^{LJ}\left(r_{i,j}\right)
+\sum\limits_{\substack{k=1 \\ k\ne i,j} }^N
{\sum\limits_{\substack{\ell=1 \\ \ell\ne i,j,k}}^N
{\varphi_{kij\ell}^{tor}}}
\end {array}
\right]} }
\end{equation}

\noindent where $\varphi^R$ and $\varphi^A$ are the interatomic repulsion and attraction terms between valence electrons, respectively, for bound atoms. The long-range interactions are included by adding $\varphi^{LJ}$, a parameterized \textit{Lennard-Jones 12-6} potential term. $\varphi^{tor}$ is a single-bond torsion term. The bond order function $b_{ij}$ includes the many body effects,

\begin{equation}
\label{eq:2}
b_{ij} = \frac{1}{2} \left( b_{ij}^{\sigma-\pi} + b_{ji}^{\sigma-\pi} + b_{ji}^{RC}+ b_{ji}^{DH} \right)
\end{equation}

\noindent where $b_{ij}^{\sigma-\pi}$ depends on the atomic distance and bond angle, $b_{ji}^{RC}$ represents the influence of bond conjugation. $b_{ji}^{DH}$ is a dihedral-angle term for double bonds. This potential is an extension of the second generation of reactive empirical bond-order (REBO) model.\cite{Brenner2002} The derivatives of $-U^p$ (force components) are analytically computed. This semi-empirical approach has recently been used in many simulation works on the structural properties of carbon nanotube (CNTs) and graphene (e.g. Refs.\onlinecite{Shenoy2008,Ni2002,Zhao2009,Wang2009d,zhaowang-07-02}). It has also been used in one of our previous studies for investigating the nonlinear elasticity of CNTs,\cite{Wang2009f} in which we obtained quantitative agreement between the AIREBO-calculated Young's modulus of CNTs and that from \textit{ab-initio} calculations. Compared to first-principle methods, an important feature of this empirical potential is its ability to deal with large systems. This is particularly important in case that the number of atoms cannot be reduced by using periodic condition. 

\begin{figure}[ht]
\centerline{\includegraphics[width=9cm]{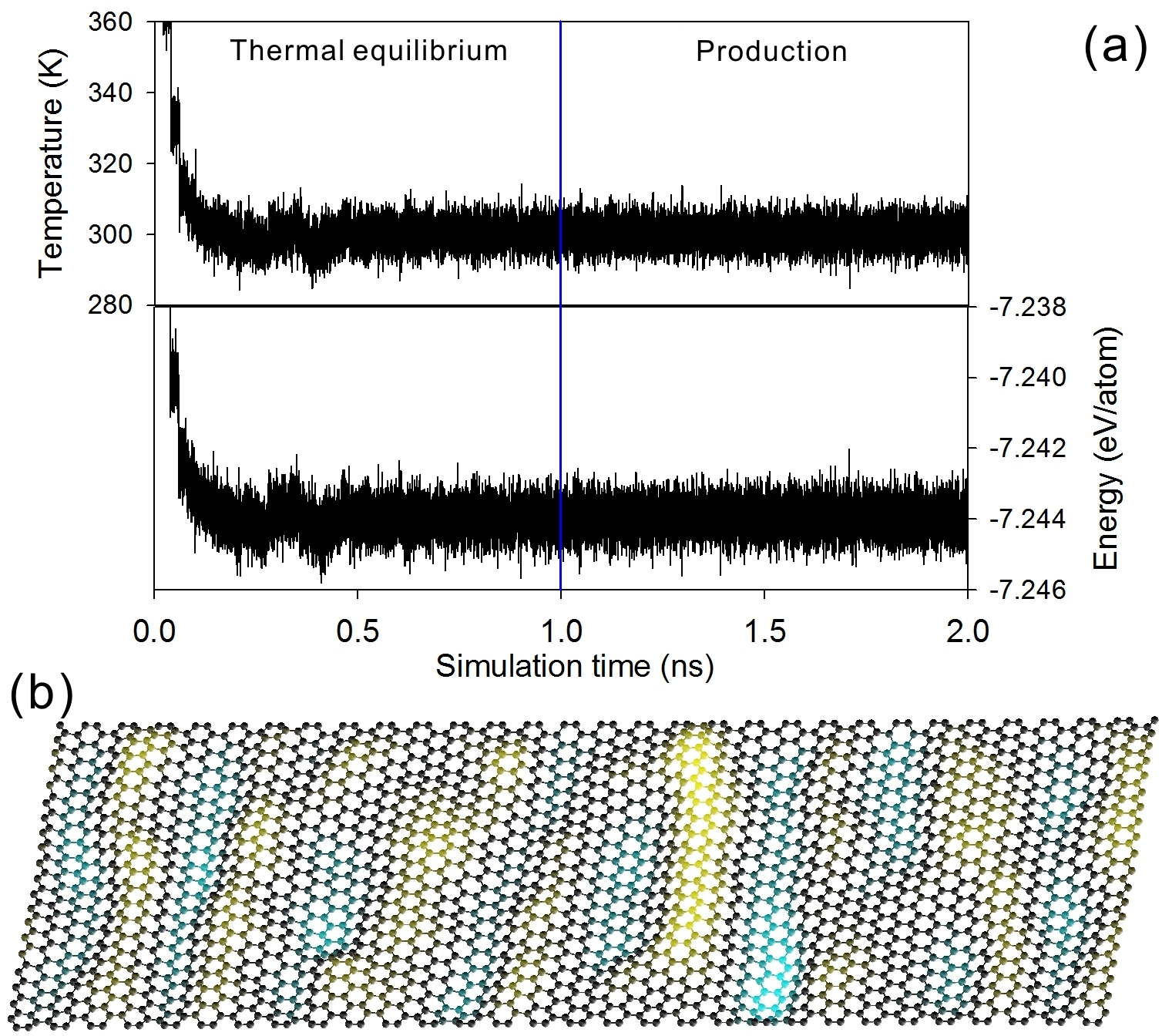}}
\caption{\label{fig:MD}
(Color online) (a) Temperature and potential energy variations during a simulation for the graphene shown in Fig.\ref{fig:Schema} (b). (b) Snapshot of the system at an arbitrary instant during the production phase.} 
\end{figure}

Fig.\ref{fig:MD}\textit{a} shows an example of how the system attains thermodynamic equilibrium during the simulation, depicting temperature and energy variations. We observe significant thermal fluctuations at the beginning of simulation. During the thermal equilibrium phase, the temperature fluctuation is reduced from $150$K to about $20$K, which corresponds to the inter-atomic potential energy variation from $20$ to $2.5$ meV/atom. The simulation result is taken as the average atomic configuration over the production phase. The resulting ripples are in well-ordered periodic wave shape (e.g. Fig.\ref{fig:Schema}\textit{b}), however, random disorder due to intrinsic thermal perturbation can be observed at any given instant during the simulation (Fig.\ref{fig:MD}\textit{b}). This is in agreement with the Monte-Carlo simulation results of Ref.\onlinecite{Fasolino2007}. 

\begin{figure}[ht]
\centerline{\includegraphics[width=9cm]{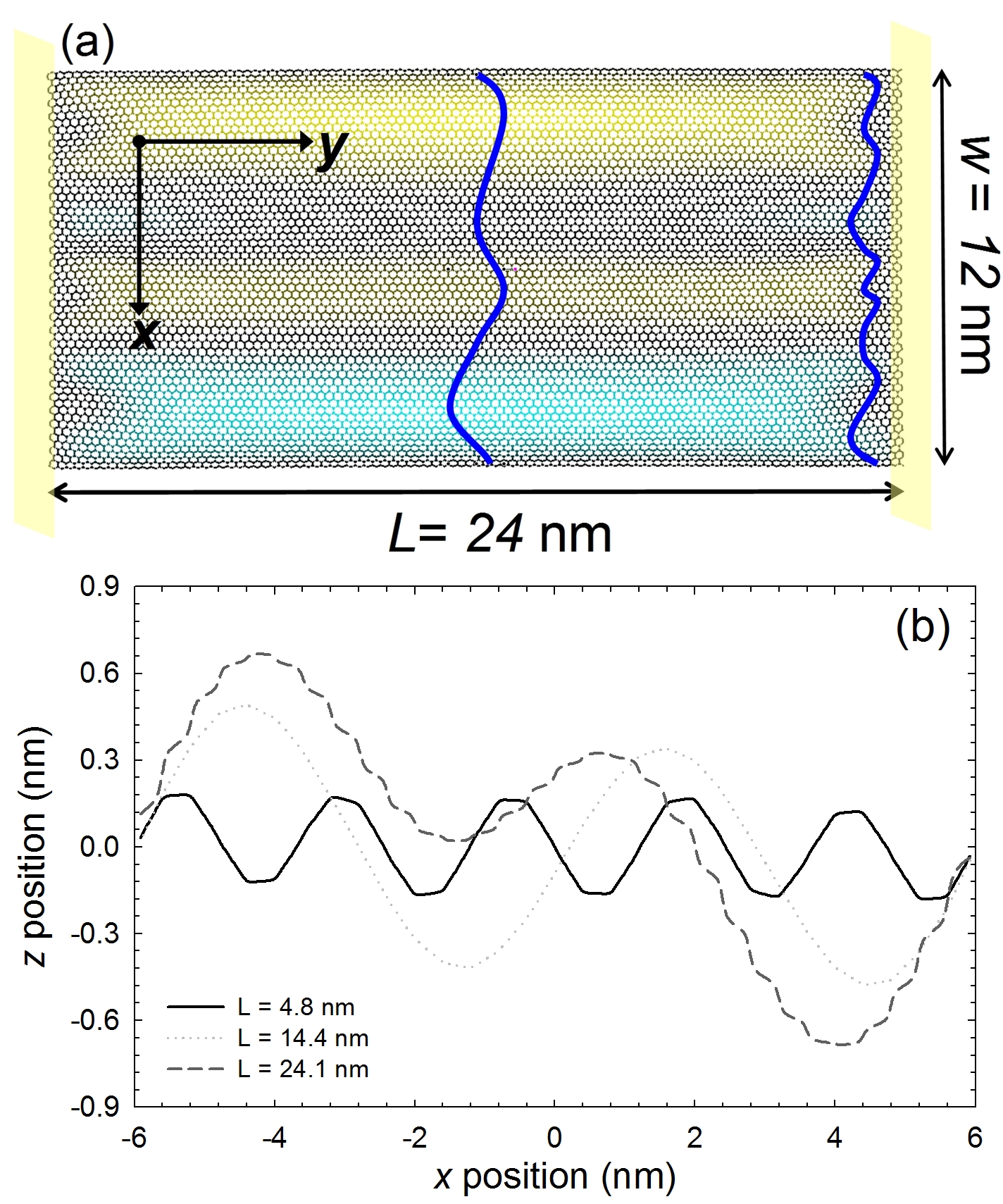}}
\caption{\label{fig:Crosssection}
(Color online) (a) Graphene ($w \times L = 12 nm \times 24 nm$) submitted to compressive edge strain $\epsilon=\Delta w / w = 0.05$. The solid lines show the shape of height profile. (b) Height profile of the middle cross section of graphene with different $L$.} 
\end{figure}

The maximum number of atoms which can be simulated by our MD code is around 20000, which is already a large number in the atomic simulation world. However, the graphene size used in experiments can be up to $10^{6}$ times larger. It is therefore important to understand the influence of graphene size on the ripple structure. In Fig.\ref{fig:Crosssection}(\textit{a}) we show the atomic configuration of a graphene sheet submitted to edge compressive strain $\epsilon=\Delta w / w=0.05$. An interesting phenomenon observed in this figure is that the waves tend to merge into each other when they propagate from the edge to the center. e.g. 6 wave undulations can be observed near the edge (solid line in Fig.\ref{fig:Crosssection}(\textit{a}), while only $2$ are left in the sheet middle (dashed line). Also, it is found that the wave amplitude in the middle is larger than that near the edges. 

\begin{figure}[ht]
\centerline{\includegraphics[width=9cm]{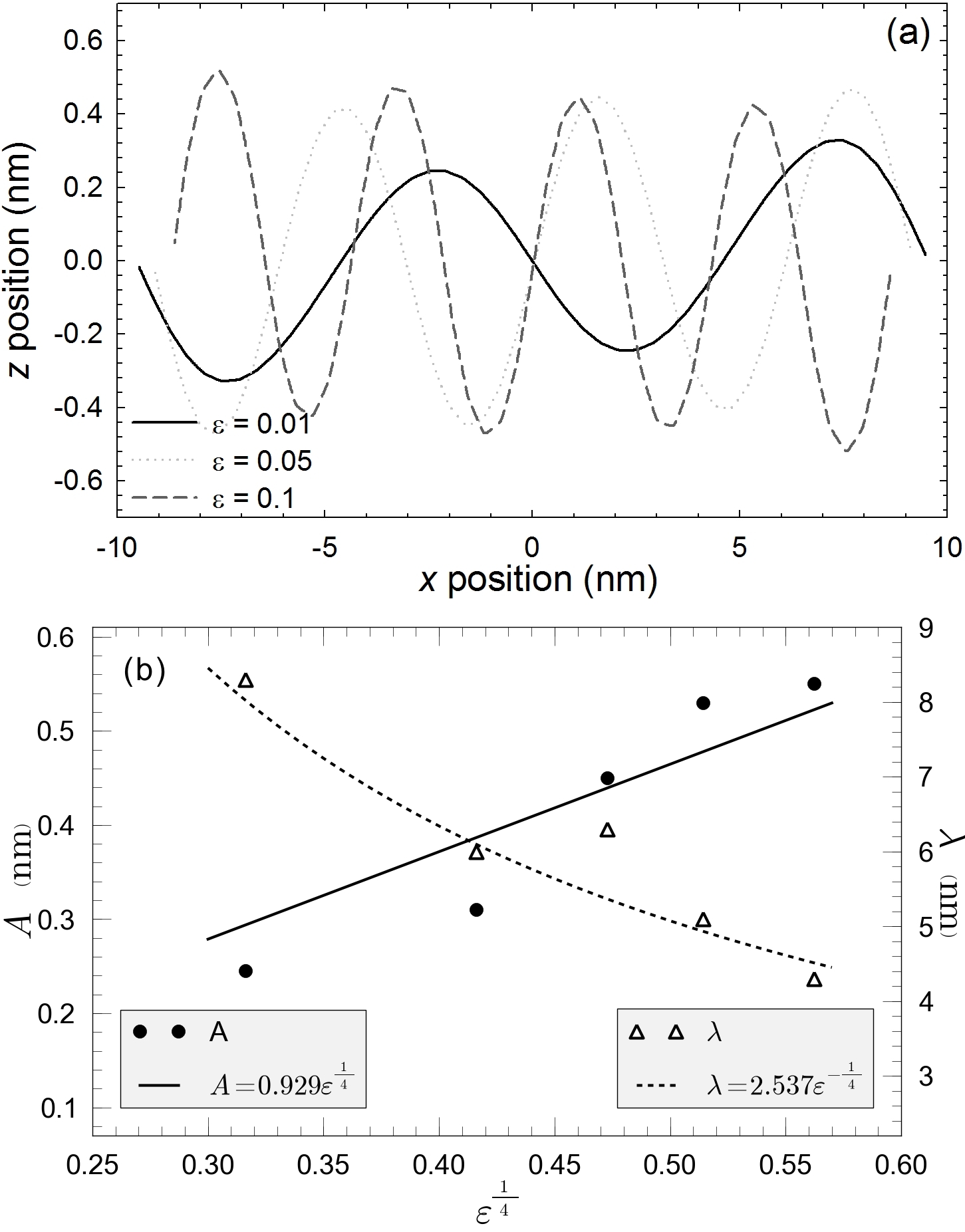}}
\caption{\label{fig:strain}
(Color online) (a) Height profile of the middle cross section of a graphene ($w \times L = 19.4 nm \times 14.6 nm$) submitted to different edge strain $\epsilon$. (b) Wavelength $\lambda$ and out-of-plane amplitude $A$ \textit{versus} $\epsilon^{1/4}$. The symbols represent simulation results and the lines stand for the best-fitted curves.} 
\end{figure}

To show the size effect, we plot in Fig.\ref{fig:Crosssection}(\textit{b}) the height profile of ripple shape at the middle of three graphene sheets with different length $L$. We can see that, for a given edge contraction, the wavelength $\lambda$ and amplitude $A$ increase with $L$. As a consequence, the ripples wave become less dense in the graphene middle and the wave form is changed. We find that the oscillation amplitude becomes about $4$ times larger for the sheet $5$ times longer. According to an analysis using the F\"{o}ppl von K\'{a}rm\'{a}n equations,\cite{Cerda2003} the wavelength and the out-of-plane displacement in the region \textit{far away} from the clamped boundaries should be roughly proportional to $\sqrt{L}$. In our simulation we observed that $A$ and $\lambda$ both increase with $L$. The exactly linear dependence on $\sqrt{L}$ is however not clearly shown, due to the fact that our graphene is not large enough. To show the full length dependence of the ripples, large-scale modeling approaches such as finite-element simulations are required for graphene size consistent with experiments, as pointed out by Shenoy \textit{et al.}.\cite{Shenoy2008}

\begin{figure}[ht]
\centerline{\includegraphics[width=9cm]{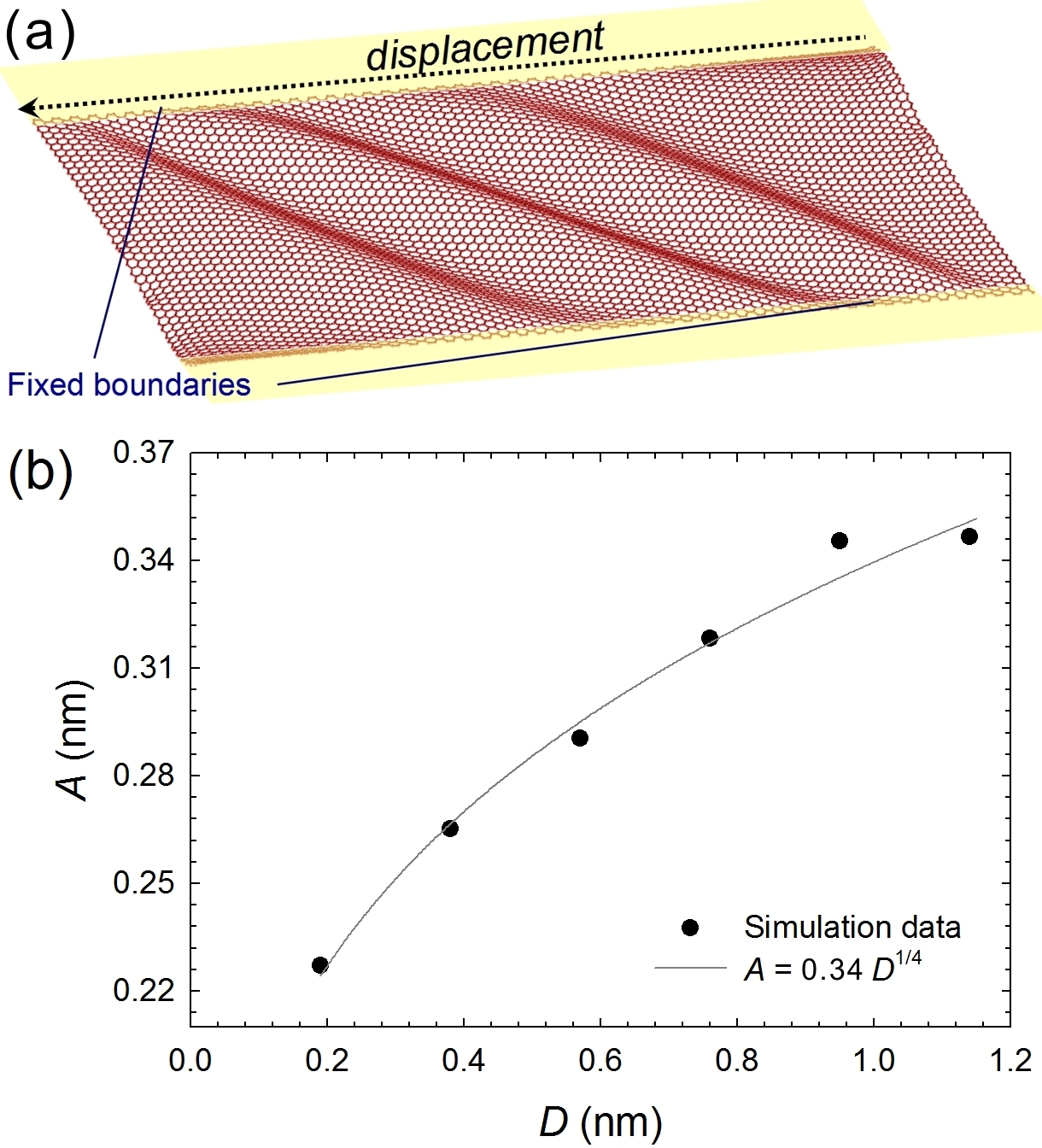}}
\caption{\label{fig:para}
(Color online) (a) Atomic configuration of a suspended graphene ($w \times L = 13 nm \times 5 nm$) submitted to $0.65$nm of edge displacement $D$ (indicated by the arrow). (b) Out-of-plane amplitude $A$ \textit{versus} $D$. The symbols represent simulation results and the curve stands for the best-fitted one. } 
\end{figure}

Fig.\ref{fig:strain}(\textit{a}) shows the height profile at the middle cross section of a graphene under compressive edge strains ranging from 0.01 to 0.1. It can be observed that more waves appear with larger oscillation amplitudes $A$ when edge strain increases. Ref.\onlinecite{Cerda2003} suggests that $\lambda^{4}$ and $A^{-4}$ should roughly hold a linear relationship with the longitudinal strain $\gamma$. Since in our simulations the ripples are induced by a lateral deformation $\epsilon$ , for establishing the correlation between the lateral strain $\epsilon$ and $\lambda$, we have used $\epsilon \approx \gamma\upsilon$ from the original definition of the Poisson's ratio $\upsilon$. Hence, according to Ref.\onlinecite{Cerda2003}, the dependence of the ripple structure on the compressive edge strain $\epsilon$ should be governed by

\begin{equation}
\label{eq:3}
\begin{array}{c}
\left\{ \begin{array}{l}
\lambda^{4} \approx {4 \pi^{2} \upsilon L^{2} t^{2}}/{\left[ 3(1-\upsilon^{2})\epsilon \right]} \\
A^{4} \approx {16\upsilon L^{2}t^{2}\epsilon}/{\left[ 3\pi^{2}(1-\upsilon^{2}) \right]}
\end{array}\right.\\
\end{array}\,,
\end{equation}

\noindent where $t$ is the thickness. The Eq.\ref{eq:3} suggests $\lambda \propto \epsilon^{-1/4}$ and $A \propto \epsilon^{1/4}$. 

These linear dependences are clearly shown when we plot the values of $\lambda$ and $A$ as functions of $\epsilon^{1/4}$. Taking $t=0.339$nm (as graphite's \textit{Van der Waals} interlayer spacing), we find that the slopes of the two best-fitted lines suggest that the value of $\upsilon$ is approximately $0.145$, which is close to the \textit{ab-initio}-calculated value.\cite{Kudin2001} However, it is necessary to note that there is still no general agreement about the value of $t$ of a one-atom-thick carbon sheet, in particular when the surface is curved and involves \textit{Van der Waals} interactions by $\pi$-stacking.\cite{Huang2006a,Vodenitcharova2003} The data fluctuation in Fig.\ref{fig:strain}(\textit{b}) is due to the small size of the simulated graphene, since the $1/4$-power law is only valid for the region far away from the clamped boundaries. In the case of thermal treatment such as that shown in Ref.\onlinecite{Bao2009}, neglecting graphene's resistance to in-plane compressive stress, the strain $\epsilon$ could be approximated by

\begin{equation}
\label{eq:4}
\epsilon \approx \Delta T(\alpha_{s} - \alpha_{g})+ \Delta T^{2}\alpha_{s}\alpha_{g}
\end{equation}

\noindent where $\Delta T$ is the magnitude of heating or cooling temperature, $\alpha_{s}$ and $\alpha_{g}$ are TEC of substrate and graphene, respectively. Neglecting the second order term, we obtain $\epsilon \approx \Delta T(\alpha_{s} - \alpha_{g})$. The wrinkling mechanism explained above also suggests that ripples can appear by only cooling without heating. 

Besides the thermal effects, our simulation results demonstrate that periodic ripples can be induced by introducing parallel displacement of the fixed boundaries. As shown in Fig.\ref{fig:para} (a), in such a case the ripples are not perpendicular to the trench, as those observed in experiments without thermal-treated samples.\cite{Chen2009} The ripple orientation seems to be not very sensitive to the magnitude of displacement $D$. The fitting curve of out-of-plane amplitude in Fig.\ref{fig:para} (b) suggests that $A$ roughly follows a 1/4-power scaling law of $D$ as $A \approx 0.34 D^{1/4}$. This correlation can be explained by the fact that $\lambda^{4}$ should be proportional to the longitudinal strain $\gamma$,\cite{Cerda2003} while the effective value of $\gamma$ should be proportional to the edge displacement $D$. This results implies that the wavelength $\lambda$ is controllable by adjusting the displacement, since $\lambda \times A $ should be roughly constant.

In summary, we have investigated the effects of edge contraction and displacement on the structural instability of suspended graphene. Our simulation results show that periodic ripples are manifested in graphene, when longitudinal compression is applied to the fixed boundaries by thermal effects. The 1/4-power law formalisms are found to be valid for the wavelength and the out-of-plane amplitude as functions of edge strain. The wavenumber is found to be larger at the graphene edge than that in the center. Besides the thermal effects, we show that parallel displacement of the fixed edge can induce periodic ripples in suspended graphene. In such a case the oscillation amplitude roughly holds a linear relationship with 1/4 power of the displacement. These results are essential for understanding the experimentally observed ripples on suspended graphene.


\end{document}